\documentclass{article}

\usepackage{style}

\usepackage[utf8]{inputenc}
\usepackage[T1]{fontenc}
\usepackage{hyperref}
\usepackage{url}
\usepackage{booktabs}
\usepackage{amsfonts}
\usepackage{amsmath}
\usepackage{nicefrac}
\usepackage{microtype}
\usepackage{paracol}
\usepackage{array}
\usepackage{lipsum}
\usepackage{fancyhdr}
\usepackage{bbm}
\usepackage{graphicx}
\usepackage{fvextra}
\DefineVerbatimEnvironment{verbatim}{Verbatim}{breaklines=true, breaksymbolleft=,fontsize=\small}

\title{Improving Embedding Accuracy for Document Retrieval Using Entity Relationship Maps and Model-Aware Contrastive Sampling}
\author{
  Thea Aviss \\
  Fifth Dimension AI \\
  London, UK\\
  \texttt{thea@fifthdimensionai.com} \\
}
\date{\today}

\begin{document}
\maketitle

\begin{abstract}
In this paper we present APEX-Embedding-7B (Advanced Processing for Epistemic eXtraction), a 7-billion parameter decoder-only text Feature Extraction Model, specifically designed for Document Retrieval-Augmented Generation (RAG) tasks. Our approach employs two training techniques that yield an emergent improvement in factual focus: (1) Pre-convergence interrupted fine-tuning using Structured Entity Relationship Maps as training data input: designed to shift the model's attention and create a bias towards factual content rather than semantic style - this enhances plain text performance despite not being directly trained for it; and (2) Model-Aware Contrastive Sampling, creating a balanced and evenly distributed collation map of hard and soft negatives directly informed by the base model's competency. This combined methodology yields significant improvements, enhancing plain text query/document pair retrieval to achieve an absolute rank@1 accuracy of 90.86\% (an increase of 6.26\% compared to the next leading model) in our evaluation, and reducing training data input context size by an average of 37.71\% compared to plain text for both queries and document texts. Based on our evaluations, our model establishes a new state-of-the-art standard in text feature extraction for longer context document retrieval tasks.
\end{abstract}

\section{Introduction}

In AI-driven document retrieval, integrating external knowledge to enhance language model responses is essential. Retrieval-Augmented Generation (RAG) improves this process by combining document retrieval with text generation, aiming for more accurate and relevant outputs \cite{lewis2021}. However, the effectiveness of RAG systems relies heavily on the quality of text embeddings used for information retrieval. Poor embeddings can introduce inaccuracies and increase the likelihood of factually incorrect or misleading responses, commonly referred to as hallucination\cite{li2024}.

To address these challenges, we present APEX-Embedding-7B, a 7-billion parameter model designed to enhance text embeddings for document retrieval tasks \cite{mistral7b,wang2024}. Our model uses structured entity relationship maps during training, focusing on key factual and relational content rather than overall semantic style. This approach ensures greater accuracy in retrieval tasks, particularly for industries with large and complex datasets, such as the property market.

For example, in property management, legal compliance, and real estate transactions, where extensive documentation is required—ranging from contracts and property deeds to market analysis reports—retrieving accurate and relevant information is critical. Our proposed model can efficiently create embeddings to allow accurate searches through vast corpora of documents, retrieving precise data such as zoning laws, transaction histories, or market trends, reducing the risk of factual errors and hallucinations. This is particularly useful when a RAG system needs to generate property listings, valuation reports, or legal advice based on detailed document archives.

By advancing text feature extraction in RAG systems, our proposed model provides a powerful solution for improving retrieval accuracy, ensuring factual precision, and streamlining information extraction in large document-heavy industries such as real estate.

\section{Proposed Methodology}
Our method for improving document retrieval accuracy in Retrieval-Augmented Generation (RAG) systems leverages two key techniques: pre-convergence interrupted fine-tuning using Structured Entity Relationship Maps and Model-Aware Contrastive Sampling. These techniques work together to enhance factual precision without sacrificing the model’s general capabilities.

\subsection{Pre-convergence Interrupted Fine-tuning with Structured Entity Relationship Maps}
This method includes fine-tuning the model with Structured Entity Relationship Maps to emphasise factual content. The structured maps are synthetic representations generated from real documents, capturing key entity relationships to shift the models attention bias toward factual data rather than linguistic style.

However, we stop the fine-tuning process before convergence for a critical reason: we want to bias the hidden state representations of the model toward factual content without diminishing its ability to work with plain text. By interrupting the process early, we prevent the model from overfitting to the structured data and becoming overly reliant on the entity maps. Instead, this approach influences the model to implicitly prioritise factual information when working with both structured inputs and plain text during general use.

The hypothesis behind our methodology can be validated by work such as Jin et al. (2024)\cite{jin2024}'s investigation of how representations of formal program semantics emerge during training. Similar to their findings that a model can learn to interpret underlying structures without explicit supervision throughout, we aim to influence our model’s internal representations toward factual embeddings. By stopping early in training, we ensure that the model maintains its general ability to handle plain text, while instilling a stronger focus on extracting critical factual details, resulting in improved embeddings for data-heavy document retrieval tasks.

In addition, recent work in parameter-efficient fine-tuning highlights the benefits of early stopping or selective updates to balance specialisation and generalisation. Houlsby et al.\cite{houlsby2022} demonstrates that limiting parameter updates in fine-tuning can prevent overfitting while still improving model performance on task-specific objectives. Similarly, Wang et al.(2022)\cite{wang2022} shows that aligning fine-tuning strategies closely with pre-training objectives can yield significant performance gains in few-shot scenarios, supporting the idea that early-stopping helps the model focus on factual accuracy while avoiding excessive reliance on structured inputs.

\subsection{Model-Aware Contrastive Sampling}
In our method, Model-Aware Contrastive Sampling improves retrieval accuracy by selecting and curating negative examples based on the base model’s existing embedding capabilities and reducing the need for large batch sizes during training. This approach primarily ensures that the negative examples used during training are tailored to the model’s strengths and weaknesses, rather than being selected arbitrarily or generated uniformly.

First, embeddings for queries and documents are generated using the base model\cite{salesforce2024}. We then compute cosine similarity\cite{luo2017} scores between these embeddings, allowing us to rank negative examples based on their similarity to the query. This ranking enables the classification of negative samples into two categories:

\textbf{Soft negatives}: These are documents with higher cosine similarity to the query but are factually incorrect or irrelevant. These examples are more challenging for the model, as they appear semantically close to the positive document but contain key factual differences.

\textbf{Hard negatives}: These are documents with lower cosine similarity to the query, and while still factually incorrect, they are easier for the model to differentiate. Introducing these examples ensures diversity in the training process and prevents overfitting to closely related examples.

We then condense a large batch of query-document pairs into smaller, well-balanced training batches consisting of one query, one positive document, five soft negatives, and five hard negatives throughout the training data and each set and distributed evenly during collation across the training data to reduce stochastic gradient noise. This balanced sampling approach helps the model improve its ability to make precise factual distinctions, which is especially important in industries like property management, where document retrieval requires high factual accuracy.

In contrast, Wang et al. 2024 \cite{wang2024} do not use the base model’s performance to inform negative sampling. Their method rests on generating synthetic data for contrastive learning tasks, where negative examples are chosen based on predefined data augmentation strategies rather than the model’s own embedding quality or performance. Their focus is on generating diverse synthetic query-document pairs for tasks like semantic similarity, with uniform sampling across these datasets. Similarly, NV-Embed \cite{lee2024} uses a separate fine-tuned encoder-based model to select hard negatives from public datasets. While both approaches rely on predefined strategies or external models for negative selection, our methodology tailors its negative sampling dynamically based on the base model’s existing capabilities, enabling it to target specific weaknesses and improve retrieval accuracy more precisely.

In the work of Huang et al. (2022)\cite{huang2022}, Model-Aware Contrastive Learning (MACL) addresses limitations in contrastive learning, such as the uniformity-tolerance dilemma, where balancing
embedding separation and semantic relationship capture can be challenging. Unlike our method, which pre-selects negatives based on similarity of embeddings created with the base model, MACL dynamically adjusts the temperature during training to regulate penalties on negatives according to the alignment of positive pairs. This adaptive strategy improves the balance between uniformity and tolerance. Additionally, MACL tackles the gradient reduction dilemma with a re-weighting strategy for efficient learning with smaller negative sample sizes, while our approach focuses on balancing hard and soft negatives to reduce batch size, reduce stochastic gradient descent noise, and improve retrieval accuracy.

\section{Model Architecture}
APEX-Embedding-7B is a decoder-only feature extraction model, based on ‘SFR-Embedding-2-R’ \cite{salesforce2024}, which itself builds on E5-mistral-7b-instruct \cite{wang2024} and Mistral-7B-v0.1 \cite{mistral7b}. Similarly, the NV-Embed-V2 model \cite{lee2024} is also based on Mistral-7B.

While both models share the same underlying base, APEX-Embedding-7B differentiates itself by leveraging structured entity relationship maps during training to improve the factual accuracy of embeddings, especially in document-heavy retrieval tasks. In contrast, NV-Embed focuses on improving general-purpose embeddings by removing the causal attention mask and introducing a latent attention layer for better sequence pooling.

Unlike NV-Embed, which modifies the attention mechanism to enhance representation, APEX-Embedding-7B retains the causal attention mask but interrupts fine-tuning pre-convergence to balance between factual emphasis and maintaining strong performance on plain text tasks. This allows APEX-Embedding-7B to focus more on structured input while still excelling in retrieval tasks that require precise factual extraction.

\section{Dataset Generation}
To train APEX-Embedding-7B, we constructed a custom dataset consisting of tens of thousands of curated document excerpts and full-text pages, drawn from a variety of publicly available sources. These documents include a mix of plain text general knowledge paragraphs, RAG corpus documents, and full legal contracts, such as property leases. Each document page is tokenised to a maximum length of 4096 tokens, matching the context window size of the base model.

Based on each processed document, query text is synthetically generated using GPT-4o\cite{gpt4o} (prompts can be found in section \ref{sec:prompts}). After this, each document and query in the training set is then represented by a Structured Entity Relationship Map, which serves as the core input format during training. These maps are also generated using GPT-4o using a specialised prompt. The maps focus on key entities and their relationships within the text, ensuring that the model learns to prioritise factual content over stylistic or irrelevant details.

The structure of an Entity Relationship Map is as follows:
\begin{verbatim}
{
    "topic": "",
    "cross_reference_tags": "",
    "entities": [{
        "entity_type": "",
        "entity_name": "",
        "attributes": {},
    }],
}
\end{verbatim}

Each resulting sample in the dataset includes the following components:

Queries: Synthetic user queries are generated based on labeled data from the corpus. These queries are meant to simulate real-world information retrieval tasks. These are represented structured as entity relationship maps.

Corpus Documents: The documents are either full pages or relevant excerpts from the corpus, represented using Structured Entity Relationship Maps. These entity maps serve as a direct transformation of the document's content, focusing on key entities and their relationships rather than stylistic or semantic features.

Relevant Document Mappings: This mapping directly connects each query to one or more relevant documents from the corpus. This is critical for training the model to retrieve the correct documents during contrastive learning.

Each sample within the dataset is structured as follows:
\begin{verbatim} 
{ "queries": { "query_id_1": "query_map_1", ... }, 
"corpus": { "document_id_1": "document_map_1", ... }, 
"relevant_docs": { "query_id_1": ["document_id_1"], ... } } \end{verbatim}

This structure supports the model’s ability to distinguish between relevant and irrelevant documents in a retrieval task, enhancing its performance in document-heavy industries like real estate, where factual precision in selecting the correct document from a pool is critical.

Our approach maintains the model’s pre-trained capability to generate high-quality plain text embeddings while enhancing its ability to handle structured entity relationship maps, as described in Section 2.1. This leads to improved factual content prioritization in document retrieval tasks.

As explained in Section 2.2, our use of Model-Aware Contrastive Sampling is central to this improvement. Both our method and Wang et al.\cite{wang2024} utilise the InfoNCE (Noise-Contrastive Estimation) loss function \cite{oord2018}. However, unlike Wang et al.\cite{wang2024}, who rely on synthetic hard negative samples generated via predefined prompting strategies, our approach selects soft and hard negatives informed by the base model’s performance. By using negative examples from the batch itself and tuning their selection based on cosine similarity\cite{luo2017} to the query, we train the model to recognise more nuanced factual distinctions, not just stark contrasts. This approach results in significantly improved embedding quality and retrieval accuracy, particularly in scenarios requiring high factual precision.

\section{Applying Model Aware Contrastive Sampling}

Our collation strategy is an essential part of the Model-Aware Contrastive Sampling methodology, designed to optimise training efficiency by reducing the memory footprint during batch processing. Rather than treating this as a distinct process, collation leverages the same precomputed embeddings and similarity-based selection of negative samples but focuses specifically on how these are organized into smaller, more memory-efficient micro-batches.

To begin, embeddings for all queries and corpus documents are precomputed using the base model, creating a comprehensive collation map that includes cosine similarity\cite{luo2017} scores, positive matches, and ranked soft and hard negatives for each query. This map is not only used for sampling hard and soft negatives, but also to inform how these samples are distributed across micro-batches.

The core of this collation strategy is to compress large, resource-heavy batches into smaller units while maintaining the diversity and challenge necessary for effective contrastive learning. By carefully balancing hard and soft negatives within each micro-batch, the training process avoids overfitting and reduces stochastic gradient noise, all while working within the constraints of limited GPU memory.

This structured distribution of challenging samples ensures that training remains both effective and efficient, with batch sizes kept minimal without sacrificing the richness of the training data. In doing so, the model is exposed to a wide variety of contrastive examples, allowing it to refine its understanding of factual distinctions, even when operating with compressed micro-batches.

Example Collation Map Entry (positive rank@1 match)

\begin{verbatim}
"Hb4ZYSq-2g0Xgg": {
    "positives": [
        "bZjhOCwdCNixvw"
    ],
    "soft_negatives": [
        "SDBU_xkhIwcz3w",
        ...x5
    ],
    "hard_negatives": [
        "zdGv_8yIIKRSaA",
        ...x5
    ],
    "positive_match_similarity": 0.93701171875,
    "positive_rank": 0
}
\end{verbatim}
\subsection{Experimental Data and Results}

We initially conducted experiments with various batch sizes using our training dataset split to evaluate the pre-trained base model and recorded the resulting accuracy and lowest positive match rank seen during evaluation.

\begin{center}
\begin{tabular}{|c|c|c|c|c|}
\hline
Batch Size & Total Correct & Total Queries & Final Accuracy & Lowest Positive Match Rank \\ \hline
32         & 4793          & 4855          & 0.9872         & 7                          \\ \hline
256        & 4599          & 4855          & 0.9473         & 60                         \\ \hline
1500       & 4203          & 4855          & 0.8657         & 295                        \\ \hline
4855       & 3859          & 4855          & 0.7949         & 1043                       \\ \hline
\end{tabular}
\end{center}
Table 1: Evaluation results of the pre-trained model with various batch sizes.

The data demonstrates that as the batch size increases with full batch sampling for negatives, the final accuracy of the pre-trained model decreases and the lowest positive match rank seen during evaluation increases. This indicates that larger batch sizes provide a better representation of the model's shortcomings and offer a higher ceiling for improvement during further training However due to GPU Memory limitations we will have to optimise this to give us the same improvement potential with much smaller batch sizes.

\subsection*{Base Model Evaluation Process\footnote{Our Evaluation process in section \ref{sec:eval} follows a very similar procedure.}}

\subsection*{Cosine Similarity}
For each query embedding \( \mathbf{q} \) and document embedding \( \mathbf{d}_i \), cosine similarity\cite{luo2017} is:

\[
\text{cosine\_similarity}(\mathbf{q}, \mathbf{d}_i) = \frac{\mathbf{q} \cdot \mathbf{d}_i}{\|\mathbf{q}\| \|\mathbf{d}_i\|}
\]
Cosine similarity\cite{luo2017} measures the cosine of the angle between two vectors in the embedding space, in this case between the query embedding \( \mathbf{q} \) and the document embedding \( \mathbf{d}_i \). It gives a value between -1 and 1, where 1 means the vectors are perfectly aligned (most similar) and -1 means they are diametrically opposed (least similar). This similarity score is essential for ranking the relevance of documents to a given query in retrieval tasks.

\subsection*{Document Ranking}
After queries and document texts are generated by the base model we perform the following.
For each query \( j \), we define a set of candidate documents, denoted as \( D_j \). We calculate the similarity between the query \( j \) and each document \( \mathbf{d}_i \) in \( D_j \) using cosine similarity:

\[
\text{similarity}_j = \{\text{cosine\_similarity}(\mathbf{q}_j, \mathbf{d}_i) \mid i \in D_j\}
\]

The documents are then sorted based on their similarity scores to produce a ranked list, \( \text{ranked\_docs}_j \). The rank of the positive (relevant) document, \( d_{\text{pos}} \), within this list is defined as:

\[
\text{positive\_rank}_j = \text{rank of } d_{\text{pos}} \text{ in } \text{ranked\_docs}_j
\]

\subsection*{Accuracy Calculation}
The accuracy is the proportion of queries where the positive document is ranked first:

\[
\text{Accuracy} = \frac{\sum_{j=1}^{n} \mathbbm{1}(\text{positive\_rank}_j = 1)}{n}
\]

Here, \( \mathbbm{1}(\text{positive\_rank}_j = 1) \) is an indicator function that returns 1 if the positive document is ranked first for query \( j \), and 0 otherwise. The total number of queries is denoted by \( n \).

\clearpage
\subsection{Determining the Smallest Effective Batch Size}

Following the completion of the Model-Aware Contrastive Sampling process, the next task is to reorganise the collation map to identify the smallest effective batch size for training. This step relies on redistributing the challenging samples (i.e., soft and hard negatives) across the batches, ensuring that each batch has a balanced proportion of positive examples while maintaining the difficulty level seen in larger batches. The goal is to retain the same level of challenge in smaller batches while optimising memory usage.

To maintain consistency across training, the aim is to ensure that each batch reflects the base model’s observed performance distribution, meaning each batch should contain a similar mix of rank 1 correct predictions and failed queries. This ensures that the model is exposed to a representative sample of both successful and challenging cases in every training step.

To determine the smallest effective batch size, we apply the following process:

\[
b_{\text{min}} = \min \left\{ b \in \mathbb{N} \mid \left(\frac{N_{\text{queries}}}{b} \geq 1\right) \land \left(\frac{N_{\text{positives}}}{\lceil \frac{N_{\text{queries}}}{b} \rceil} \geq 1\right) \right\}
\]

Where:
\begin{itemize}
    \item \( N_{\text{queries}} \): Total number of queries in the dataset.
    \item \( N_{\text{positives}} \): Total number of queries where the positive document is ranked at 0 (i.e., \( \text{positive\_rank} = 0 \) or rank@1 if non-zero indexed).
    \item \( \lceil \frac{N_{\text{queries}}}{b} \rceil \): Number of batches for a given batch size \( b \).
    \item \( b_{\text{min}} \): The smallest batch size \( b \) that guarantees a balanced distribution of positive ranks across all batches.
\end{itemize}

\subsection*{Number of Batches and Positives per Batch}
Once the smallest batch size \( b_{\text{min}} \) is found, the total number of batches is:

\[
\text{num\_batches} = \left\lceil \frac{N_{\text{queries}}}{b_{\text{min}}} \right\rceil
\]

The number of queries with \( \text{positive\_rank} = 0 \) (i.e., positive matches) per batch is:

\[
\text{pos\_per\_batch} = \max \left( 1, \left\lfloor \frac{N_{\text{positives}}}{\text{num\_batches}} \right\rfloor \right)
\]

\subsection*{Percentage of Positive Matches per Batch}
The percentage of queries with \( \text{positive\_rank} = 0 \) (i.e., positive matches rank@1) in each batch \( i \) is:

\[
\text{Percentage}_{i} = \frac{\text{Positive Matches in Batch } i}{\text{Total Queries in Batch } i} \times 100
\]

The challenging samples are distributed evenly across batches to ensure that each batch maintains the same level of difficulty as observed in larger batches. This approach guarantees that each batch contains a balanced mix of positive and negative samples, which promotes more robust learning, reduces the risk of over-fitting, and minimises stochastic gradient noise—issues that typically arise in smaller, unbalanced batches. This method effectively condenses the learning effectiveness of very large batches into much smaller, memory-efficient batches.

This collation strategy presents an effective approach to training with smaller batch sizes by embedding the model's pre-training performance into the training process. By ensuring each batch is both challenging and representative, the strategy enhances training efficiency while reducing the memory footprint. This enables the training of large-scale models with more efficient resource usage. The balanced exposure to diverse data points helps improve the model's accuracy and generalisation, contributing to its ability to handle complex queries with limited computational resources.
\clearpage 

\section{Training}
For training, We employed 4-bit QLoRA (Quantised Low-Rank Adaptation) techniques \cite{dettmers2023} to enable efficient fine-tuning while preserving the integrity of the pre-trained weights. This approach allowed for resource-efficient training by modifying only a small number of parameters, reducing the computational burden without compromising the model’s capacity. Our custom training loop was implemented using the PEFT (Parameter-Efficient Fine-Tuning) \cite{peft2024} and Transformers \cite{hftransformers} libraries.

The training process was configured with the following hyperparameters and loss function:

\begin{itemize}
    \item Effective Batch Size: 10
    \item Gradient Accumulation Steps: 2
    \item Loss Function: Noise-Contrastive Estimation (InfoNCE) with a temperature of 0.07
    \item Learning Rate: 2e-5
    \item LoRA Configuration:
    \begin{itemize}
        \item Rank (\(r\)): 8
        \item Alpha: 16
        \item Applied to all linear layers: \texttt{q\_proj}, \texttt{k\_proj}, \texttt{v\_proj}, \texttt{o\_proj}, \texttt{gate\_proj}, \texttt{up\_proj}, \texttt{down\_proj}
    \end{itemize}
\end{itemize}

The training ran for 545 steps, with the CUDA memory reserved maxing out at 38GB. The total training time was approximately 13 hours on a single 40GB NVIDIA A100 GPU.

Validation was conducted every 100 steps to monitor model performance and ensure the model did not overfit to the structured entity relationship maps. The training process was stopped early after peak accuracy was reached, as further training risked degrading plain text performance due to overfitting on the structured data.

A key aspect of the training duration was the frequent validation checks, which significantly extended the training time. If the validation process had been optimized, the total training time could have been reduced.

In contrast to our pre-convergence interrupted fine-tuning, which halts early to balance factual accuracy and generalisation, the NV-Embed model \cite{lee2024}, employs a two-stage contrastive instruction-tuning process. The first stage focuses specifically on retrieval tasks, utilising in-batch negatives and hard negatives generated by a separate  encoder based model to refine the model’s ability to differentiate between closely related documents. In the second stage, NV-Embed is fine-tuned on a broader range of tasks, including classification and clustering, without the use of in-batch negatives, to enhance its generalization across multiple language domains. Additionally, while our model applies the same instruction across both queries and documents to streamline training and improve factual precision, NV-Embed adapts different instructions for retrieval and non-retrieval tasks. This choice allows NV-Embed to handle diverse language tasks but may trade off retrieval-specific performance optimization, where our model remains focused.

\subsection{Observations on Convergence}
One notable observation is that if the model were allowed to fully converge on the structured entity relationship maps, the improvement in evaluation accuracy using a strucural entity map version of our evaluation dataset would be marginal, only exceeding the base model’s plain text accuracy by 1.93\%. The base model achieved 84.6\% accuracy on plain text, whereas the model trained to convergence on the entity relationship maps improved this to 86.53\%.

This outcome helps to further validate our methodology and hypothesis that interrupting training before full convergence on entity relationship maps refocuses the model on key factual differentiators\cite{houlsby2022}, rather than solely optimising for semantic similarities. By halting training early, we allow the model to generalise better to plain text while still benefiting from the structured data, effectively balancing the model’s ability to handle factual content across different types of input.

\clearpage
\section{Evaluation}
\label{sec:eval}
\subsection{Evaluation Methodology}

In evaluating our embedding model and other leading embedding models for comparison, we aimed to ensure that the evaluation methodology reflected real-world use within a Retrieval-Augmented Generation (RAG) application as closely as possible.

\subsection{Dataset}

Our evaluation dataset consists of 1,500 previously selected queries and their respective documents, unseen to our model and split before training. These documents have a context length spread representative of the training dataset ($\leq$4096 tokens). The documents cover a wide range of topics, including: General knowledge, Legal texts, Property market-related content (e.g., leases and contracts). For each query and document, embeddings were generated iteratively via last token pooling, with the exception of proprietary models from OpenAI and Google, which were accessed via API.

\subsection{Evaluation Criteria}

In an accurate RAG system, it is crucial that the top-ranked retrieved document is the correct one, to a) Prevent factual inaccuracies and potential misinformation in responses, and b) Reduce the likelihood of hallucination if the document text is passed to a Large Language Model (LLM). Therefore our evaluation methodology centers on testing the quality and factual alignment of the vector embeddings within a rigorous framework.

Our evaluation process is both comprehensive and exacting. For each query in our dataset, we compare its embedding against all 1,500 document embeddings in the evaluation dataset. This exhaustive approach ensures that no challenging samples are overlooked in our analysis. The comparison between each query embedding and all document embeddings is carried out using cosine similarity \cite{luo2017}, a metric chosen for its proven efficacy in exhaustively measuring the similarity between high-dimensional vectors \cite{luo2017}.
Following the calculation of similarities, we sort the entire list of documents in descending order of similarity scores for each query \cite{luo2017}. We then scrutinise the document at rank 1 (highest similarity) to ascertain whether it corresponds to the correct document for the given query. Our criteria are stringent: if the top-ranked document is indeed the correct match, we assign a score of 1; otherwise, it receives a score of 0. This binary scoring system reflects our emphasis on precise, top-ranking retrieval.
To obtain our final metric, we calculate an average accuracy across the entire dataset. This is achieved by summing the binary scores and dividing by the total number of queries, yielding a comprehensive measure of our model's performance.

This approach ensures that our evaluation closely mirrors the performance requirements of real-world RAG applications.

\subsection{Results}
\begin{center}
\begin{tabular}{>{\raggedright\arraybackslash}p{0.22\textwidth}>{\raggedright\arraybackslash}p{0.28\textwidth}>{\centering\arraybackslash}p{0.15\textwidth}>{\raggedleft\arraybackslash}p{0.15\textwidth}}
\toprule
\textbf{Manufacturer} & \textbf{Model} & \textbf{Accuracy (\%)} & \textbf{MTEB\cite{mteb} Rank*} \\
\midrule
\textbf{Fifth Dimension AI} & \textbf{APEX-Embedding-7B} & \textbf{90.86} & - \\
Salesforce & SFR\_Embedding\_2-R & 84.60 & 6th \\
OpenAI & text-embedding-3-large & 79.13 & 28th \\
Google & text-embedding-004 & 79.00 & 25th \\
BAAI & BGE-EN-ICL & 76.86 & 2nd \\
NVIDIA & nv-embed-v2 & 71.87 & 1st \\
Dunzhang & Stella\_en\_1.5B\_V5 & 69.86 & 3rd \\
\bottomrule
\end{tabular}
\end{center}
* Retrieval Task Rank Only (as of September 2024).

Table 2: Evaluation of Retrieval Accuracy (@ rank 1). Showing that our model can be regarded as the now current SoTA (State of the Art) for this task.

It's important to note that our model has not yet been evaluated on the retrieval tasks within MTEB (Massive Text Embedding Benchmark)\cite{mteb}. That being said, our results provide some evidence to show that the Retrieval-specific subset of MTEB may not accurately reflect real-world use cases, particularly when it comes to retrieving longer context documents. This discrepancy highlights the need for benchmarks that better represent the challenges of retrieving information from extensive, context-rich documents.

\clearpage

\section{Conclusion}

In this paper, we presented APEX-Embedding-7B, a language model optimised for document retrieval in RAG. By integrating pre-convergence interrupted fine-tuning with Structured Entity Relationship Maps and Model-Aware Contrastive Sampling, we enhanced both factual precision and training efficiency. These methods allowed us to focus the model on critical content without over-fitting, and efficiently manage training with reduced memory usage.

APEX-Embedding-7B has demonstrated a significant improvement in rank@1 retrieval accuracy, surpassing existing state-of-the-art models by 6.26\% (achieving 90.86\%), while also reducing input context size by 37.71\% during training. These results highlight its potential to enhance retrieval accuracy in industries like property management and legal services, where factual accuracy is crucial and also demonstrates practical steps to improve training efficiency.

Future work could explore scaling these techniques for larger models and expanding their application to other domains, as well as optimising training processes further for even greater efficiency. The model's context window can also be expanded with techniques such as RoPE scaling and rotation  \cite{su2021} (section 5.2 \cite{wang2024}). We believe our approach offers a promising path forward for improving document retrieval systems that require both high precision and practical scalability.

\section{Appendix}
\subsection{Prompts Used For Dataset Generation Only}
\label{sec:prompts}
Query Generation\footnote{Including "versioning" in prompts, such as "v2.0," can enhance LLM output quality by leveraging the model's learned associations from its training data, where higher version numbers typically indicate improvements. This approach aligns with research in prompt engineering, which demonstrates that clearly defined tasks and contextual cues, like dates or phases, can improve model performance. For instance, models have been shown to adjust their outputs based on subtle contextual shifts, producing more relevant results when prompted with cues tied to progression or refinement \cite{prompt_report, rope_study}. By referencing "v2.0" in a prompt, the model might infer a requirement for higher-quality output, drawing on patterns where version numbers are associated with enhanced features \cite{dtg_study}.}:
\begin{verbatim}
""""
User Query Generator v2.0
Purpose: Expertly analyze the text excerpt given and output a query or question sentence that includes references to the task at hand and type of document:\n\n
Please fulfill this purpose with precision and factual accuracy. You must write the query without any "document name"... It should be framed as a question, with detail.
"""
\end{verbatim}

Structured Entity Relationship Map
\begin{verbatim}
"""RAG Metadata Preparation System v2.0\n\n
Purpose: Expertly analyze the text excerpt given and output the following JSON:\n\n
{
  "topic": "",
  "cross_reference_tags": "",
  "entities": [{
    "entity_type": "",
    "entity_name": "",
    "attributes": {}, 
  }],
}

There must be no detail lost or omitted, and please intracately capture the factual structure of the entity relationships within. Details that would otherwise be lost in a general "semantic" understanding must be highlighted here.

Please fulfill this purpose with precision and factual accuracy.
"""
\end{verbatim}
After this, all generations undergone human quality control and editing.

\subsection{Feature Extraction Prompts}
\label{sec:fe_prompts}
This instruction is used for both queries and documents/passages during end-use of the model.
\begin{verbatim}
"Instruction: Please perform a RAG search based on the following. Text: {query_or_passage}"
\end{verbatim}

\end{document}